\documentclass[12pt]{article}
\usepackage{a4wide}
\usepackage{latexsym}
\usepackage{amsmath}
\usepackage{amsfonts}
\usepackage{amscd}
\usepackage{cite}
\usepackage{graphicx}
\usepackage{axodraw}
\usepackage{float}

\usepackage{pslatex}
\usepackage[latin1]{inputenc}
\usepackage[OT2,T1]{fontenc}

\newcommand{\bq}{\begin{eqnarray}}
\newcommand{\eq}{\end{eqnarray}}
\newcommand{\eps}{\varepsilon}

\DeclareSymbolFont{cyrletters}{OT2}{wncyr}{m}{n}
\DeclareMathSymbol{\Sha}{\mathalpha}{cyrletters}{"58}

\begin{document}

\thispagestyle{empty}

\begin{flushright}
  MITP/15-059
% \\ version of \today
\end{flushright}

\vspace{1.5cm}

\begin{center}
  {\Large\bf Proof of the fundamental BCJ relations for QCD amplitudes\\
  }
  \vspace{1cm}
  {\large Leonardo de la Cruz, Alexander Kniss and Stefan Weinzierl\\
\vspace{2mm}
      {\small \em PRISMA Cluster of Excellence, Institut f{\"u}r Physik, }\\
      {\small \em Johannes Gutenberg-Universit{\"a}t Mainz,}\\
      {\small \em D - 55099 Mainz, Germany}\\
  } 
\end{center}

\vspace{2cm}

% abstract -------------------------------------------------------------------------
\begin{abstract}\noindent
  {
The fundamental BCJ-relation is a linear relation between primitive tree amplitudes with different 
cyclic orderings. 
The cyclic orderings differ by the insertion place of one gluon.
The coefficients of the fundamental BCJ-relation are linear in the Lorentz invariants $2 p_i p_j$.
The BCJ-relations are well established for pure gluonic amplitudes as well as
for amplitudes in ${\mathcal N}=4$ super-Yang-Mills theory.
Recently, it has been conjectured that the BCJ-relations hold also for QCD amplitudes.
In this paper we give a proof of this conjecture.
The proof is valid for massless and massive quarks.
   }
\end{abstract}

\vspace*{\fill}

% main text ------------------------------------------------------------------------
\newpage

% ----------------------------------------------------------------------------------
\section{Introduction}
\label{sect:intro}

Amplitudes in QCD are often computed by a decomposition into a sum of smaller pieces, called primitive
amplitudes \cite{Bern:1994fz,Reuschle:2013qna}.
The primitive amplitudes are gauge invariant, colour-stripped and have a fixed ordering of the external legs.
Non-trivial relations between different primitive tree amplitudes are a fascinating topic 
and have important applications.
For pure gluonic primitive tree amplitudes these relations are by now well-studied.
Invariance under cyclic permutations is trivial. The first non-trivial relations are
the Kleiss-Kuijf relations \cite{Kleiss:1988ne}, which follow from the anti-symmetry of the colour-stripped vertices.
More interesting are the Bern-Carrasco-Johansson relations (BCJ-relations) \cite{Bern:2008qj}.
The full set of the BCJ-relations follows from the so-called fundamental BCJ-relations \cite{Feng:2010my}.
The fundamental BCJ-relation is a linear relation between primitive tree amplitudes with different cyclic orderings. 
The cyclic orderings differ by the insertion place of one gluon.
In the fundamental BCJ-relation the coefficients of the relation are linear in the Lorentz invariants $2 p_i p_j$.
The BCJ-relations are known to hold for pure gluonic tree amplitudes and for tree amplitudes in ${\mathcal N}=4$
super-Yang-Mills theory.
The BCJ relations have been proven first with methods from 
string theory \cite{BjerrumBohr:2009rd,Stieberger:2009hq}
and later within quantum field theory with the help of on-shell recursion relations \cite{Feng:2010my,Jia:2010nz,Chen:2011jxa}.
On-shell recursion relations require a certain fall-off behaviour for large momentum deformations.
For amplitudes in ${\mathcal N}=4$ SYM it is essential that not only the (bosonic) momentum components but also the 
(fermionic) Grassmann components are shifted.
The required fall-off behaviour has been established for pure gluonic tree amplitudes 
and amplitudes in ${\mathcal N}=4$ SYM in \cite{ArkaniHamed:2008yf,ArkaniHamed:2008gz,Cheung:2008dn}.
Furthermore BCJ relations have been derived for a pair of massive scalars 
and an arbitrary number of gluons \cite{Naculich:2014naa}.

It is natural to consider primitive tree amplitudes in QCD as well, involving in addition to gluons
massless and/or massive quarks \cite{Johansson:2015oia,Mastrolia:2015maa}.
The fundamental BCJ-relation singles out three particles, which we will label $1$, $2$ and $n$.
In the fundamental BCJ-relation the positions of particles $1$ and $n$ are fixed, as there the positions of the remaining particles
$3$ to $(n-1)$. In the cyclic order particles $1$ and $n$ are adjacent.
In the fundamental BCJ-relation particle $2$ is inserted in all possible places in the cyclic order between $1$ and $n$, but not between $n$ and $1$. 
Recently, Johansson and Ochirov conjectured \cite{Johansson:2015oia} 
that the fundamental BCJ-relations hold for primitive tree amplitudes in full QCD as well, provided particle $2$ is a gluon.
In this paper we prove this conjecture.
The major part of our arguments is not specific to four space-time dimensions.
Only in the explicit definitions of the momentum shifts we use spinor representations 
corresponding to four space-time dimensions.
With a suitable generalisation of these momentum shifts our proof will carry over to $D$ space-time dimensions.

This paper is organised as follows:
In section~(\ref{sect:overview}) we introduce the conjecture, which we are going to prove and give an outline
of the proof.
Section~(\ref{sect:deformation}) carefully defines momentum deformations through three-particle shifts.
This is a necessary technical preparation for our proof.
In section~(\ref{sect:large_z_behaviour}) we discuss the large $z$-behaviour of the deformed fundamental BCJ-relation
under the three-particle shifts and show that there is no contribution from infinity in BCFW-recursion relations.
In section~(\ref{sect:induction}) we prove the fundamental BCJ-relation by induction.
Our conclusions are given in section~(\ref{sect:conclusions}).
In an appendix we collected some technical details on certain three-particle shifts with massive quarks (appendix~\ref{appendix:existence_solution})
and a proof on the large $z$-behaviour in the eikonal approximation (appendix~\ref{appendix:eikonal}). 
For the convenience of the reader we also included the cyclic-ordered Feynman rules relevant to primitive amplitudes in appendix~\ref{appendix:cyclic_ordered_rules}.

% ----------------------------------------------------------------------------------
\section{Overview}
\label{sect:overview}

Tree amplitudes in QCD with any number of external quarks 
can be decomposed systematically into gauge-invariant primitive amplitudes 
with a fixed cyclic ordering \cite{Mangano:1991by,Reuschle:2013qna,Schuster:2013aya}.
Let us consider a tree-level primitive QCD amplitude 
with $n$ external particles, 
out of which $n_q$ particles are quarks, $n_q$ particles are anti-quarks and $n_g$ particles are gluons. 
We have the obvious relation
\bq
 n & = & n_g + 2 n_q.
\eq
Without loss of generality we may assume that all quarks have different flavours.
The quarks may be massless or massive.
In this paper we consider amplitudes with at least one gluon ($n_g>0$).
This excludes the case, where all external particles are either quarks or anti-quarks ($n_g=0$ and hence $n=2n_q$).
We are mainly interested in the case, where there is in addition to gluons at least one quark-anti-quark pair ($n_q>0$).
The pure gluonic case ($n_q=0$) is not excluded, but already well studied.
The tree-level primitive QCD amplitudes are cyclic-ordered.
We denote such an amplitude with the cyclic-order $(1,2,...,n)$ by
\bq
 A_n\left(1,2,...,n\right).
\eq
The amplitudes satisfy the Kleiss-Kuijf relations \cite{Kleiss:1988ne}.
In order to state the Kleiss-Kuijf relation
let
\bq
 \vec{\alpha} = \left( \alpha_1, ..., \alpha_j \right),
 & & 
 \vec{\beta} = \left( \beta_1, ..., \beta_{n-2-j} \right)
\eq
and $\vec{\beta}^T = ( \beta_{n-2-j}, ..., \beta_1 )$.
The Kleiss-Kuijf relations read
\bq
\label{Kleiss_Kuijf}
 A_n\left( 1, \vec{\beta}, 2, \vec{\alpha} \right)
 & = & 
 \left( -1 \right)^{n-2-j}
 \sum\limits_{\sigma \in \vec{\alpha} \; \Sha \; \vec{\beta}^T}
 A_n\left( 1, 2, \sigma_1, ..., \sigma_{n-2} \right).
\eq
Here, $\vec{\alpha} \; \Sha \; \vec{\beta}^T$ denotes the set of all shuffles of $\vec{\alpha}$ with $\vec{\beta}^T$, 
i.e. the set of all permutations of the elements of $\vec{\alpha}$ and $\vec{\beta}^T$, 
which preserve the relative order of the elements of $\vec{\alpha}$ and of the elements of $\vec{\beta}^T$.
A special case of the Kleiss-Kuijf relations is the situation, where the set $\beta$ contains only one element. 
In this case the Kleiss-Kuijf relation reduces to the $U(1)$-decoupling identity
\bq
\label{basic_Kleiss_Kuijf}
 \sum\limits_{\sigma \in {\mathbb Z}_{n-1}}
 A_n\left(\sigma_1,\sigma_2,...,\sigma_{n-1},n\right) 
 & = & 0,
\eq
where the sum is over the cyclic permutations of the first $(n-1)$ arguments.
The Kleiss-Kuijf relations in eq.~(\ref{Kleiss_Kuijf}) allow us to fix two legs at specified positions.
We will use this freedom to fix a particle at position $1$ and a second particle at position $n$.
Since we also assume that there is at least one gluon, let us label this gluon by $2_g$.
In \cite{Johansson:2015oia} Johansson and Ochirov conjecture that
\bq
\label{fundamental_BCJ_relation}
 \sum\limits_{i=2}^{n-1} 
  \left( \sum\limits_{j=i+1}^n 2 p_2 p_j \right)
  A_n\left(1,3,...,i,2_g,i+1,...,n-1,n\right)
 & = & 0.
\eq
In this paper we prove this conjecture.
An equivalent formulation of eq.~(\ref{fundamental_BCJ_relation}) is
\bq
\label{fundamental_BCJ_relation_2}
 \sum\limits_{i=2}^{n-1}
  \left( \sum\limits_{j=1}^i 2 p_2 p_j \right)
  A_n\left(1,3,...,i,2_g,i+1,...,n-1,n\right)
 & = & 0,
\eq
which follows from eq.~(\ref{fundamental_BCJ_relation}) by momentum conservation.
Eq.~(\ref{fundamental_BCJ_relation}) is the fundamental BCJ-relation \cite{Bern:2008qj,Feng:2010my}
for tree-level primitive QCD amplitudes.
It is well known that eq.~(\ref{fundamental_BCJ_relation}) holds in the pure gluonic case.
It is also know that eq.~(\ref{fundamental_BCJ_relation}) holds for amplitudes with one quark-anti-quark pair ($n_q=1$) in massless QCD.
This follows from the fact that these amplitudes are identical to the corresponding amplitudes in ${\mathcal N}=4$ super-Yang-Mills theory,
and the fact that the BCJ-relations hold for the 
latter \cite{ArkaniHamed:2008gz,Jia:2010nz} \footnote{It is worth noting that all tree amplitudes in massless QCD can be obtained from 
tree amplitudes in ${\mathcal N}=4$ super-Yang-Mills theory \cite{Dixon:2010ik,Melia:2013bta,Melia:2013epa}.}. 
The aim of this paper is to show that eq.~(\ref{fundamental_BCJ_relation}) holds more generally in (massless or massive) QCD.

In order to prove the fundamental BCJ-relation we will make use 
of on-shell recursion relations (or Britto-Cachazo-Feng-Witten recursion relations, BCFW-recursion relations for short) \cite{Britto:2005fq}.
Within on-shell recursion relations some distinguished external momenta are deformed,
such that momentum conservation and the on-shell conditions are respected.
Let us consider a deformation of the momenta $p_1$, $p_2$ and $p_n$ depending on a variable $z$.
This is called a three-particle BCFW-shift \cite{Risager:2005vk}. 
Since for our problem there are three distinguished particles $1$, $2_g$ and $n$,
a three-particle shift is more natural as compared to a more conventional two-particle shift.
It will turn out that a three-particle shift will simplify the proof.
We denote the deformed momenta by
\bq
 \hat{p}_1(z), 
 \;\;\;\;\;\;
 \hat{p}_2(z), 
 \;\;\;\;\;\;
 \hat{p}_n(z).
\eq
We further require
\bq
 \hat{p}_1(0) = p_1, 
 \;\;\;\;\;\;
 \hat{p}_2(0) = p_2, 
 \;\;\;\;\;\;
 \hat{p}_n(0) = p_n.
\eq
For $j \neq 1,2,n$ we simply set $\hat{p}_j(z)=p_j$.
We introduce the quantity
\bq
\label{I_n}
 I_n\left(z\right)
 & = &
 \sum\limits_{i=2}^{n-1}
  \left( \sum\limits_{j=i+1}^n 2 \hat{p}_2 \hat{p}_j \right)
  A_n\left(\hat{1},3,...,i,\hat{2}_g,i+1,...,n-1,\hat{n}\right).
\eq
For $z=0$ the expression $I_n(z)$ reduces to the left-hand-side of eq.~(\ref{fundamental_BCJ_relation}).
$I_n(z)$ is clearly a rational function of $z$.
We have to show
\bq
\label{eq_to_prove}
 I_n\left(0\right) & = & 0,
\eq
or equivalently
\bq
\label{eq_to_prove_2}
 \frac{1}{2\pi i}
 \oint\limits_{z=0} \frac{dz}{z} I_n\left(z\right) & = & 0.
\eq
We do this in two steps:
We first show that $I_n(z)$ has no pole at $z=\infty$, or equivalently, that $I_n(z)$ falls off for large $z$ at least with
$1/z$.
In the second step we use induction and BCFW-recursion in order to prove eq.~(\ref{eq_to_prove}).

% ----------------------------------------------------------------------------------
\section{Momentum deformation}
\label{sect:deformation}

In this section we review the three-particle BCFW-shift for the external particles $1$, $2_g$ and $n$.
The types of particles $1$ and $n$ may either be a quark, an anti-quark or a gluon.
However, particle $2_g$ is required to be a gluon.
Quarks and anti-quarks may be massive or massless.
Therefore we will treat the general case with arbitrary masses for particles $1$ and $n$.
The case of massless quarks is included as the special case $m_1=m_n=0$.
This section may appear at a first reading a little bit technical, 
but actually it will pay off: 
We define the momentum deformations in such a way, that we obtain the same large $z$-behaviour independently of the
helicity configuration and independently of the masses of the particles.
Most of the technical details are related to massive quarks and it might be
advantageous at a first reading to pay attention to the massless case only.
In the massless case the formulae simplify considerably.

\subsection{Spinor definitions}

For the definition of massive spinors we follow \cite{Schwinn:2007ee}.
Let us consider two independent Weyl spinors $|q+\rangle$ and $\langle q+|$. These two Weyl spinors define
a light-like four-vector
\bq
 q^\mu & = & \frac{1}{2} \langle q+ | \gamma^\mu | q+ \rangle.
\eq
This four-vector can be used to associate to any not necessarily light-like four-vector $p$ a light-like
four-vector $p^\flat$:
\bq
\label{projection_null}
 p^\flat & = & p - \frac{p^2}{2 p \cdot q} q.
\eq
The four-vector $p^\flat$ satisfies $(p^\flat)^2=0$.
Note that $p^\flat$ depends implicitly on $|q+\rangle$ and $\langle q+|$.
The two Weyl spinors $|q+\rangle$ and $\langle q+|$ are used as reference spinors in the definition of the polarisations of the 
external particles. 
For quarks with momentum $p$ we take the spinors $u$ and $\bar{u}$ as
\bq
\label{def_massive_spinors_u} 
 u(-) = \frac{1}{\langle p^\flat + | q - \rangle}  \left( p\!\!\!/ + m \right) | q - \rangle,
 & &
\bar{u}(+) = \frac{1}{\langle q - | p^\flat + \rangle} \langle q - | \left( p\!\!\!/ + m \right),
 \nonumber \\
 u(+) = \frac{1}{\langle p^\flat - | q + \rangle} \left( p\!\!\!/ + m \right) | q + \rangle,
 & &
\bar{u}(-) = \frac{1}{\langle q + | p^\flat - \rangle} \langle q + | \left( p\!\!\!/ + m \right).
\eq
The spinors $v$ and $\bar{v}$ are given by
\bq
 v(-) = \frac{1}{\langle p^\flat + | q - \rangle} \left( p\!\!\!/ - m \right) | q - \rangle,
 & &
 \bar{v}(+) = \frac{1}{\langle q - | p^\flat + \rangle} \langle q - | \left( p\!\!\!/ - m \right),
 \nonumber \\
 v(+) = \frac{1}{\langle p^\flat - | q + \rangle} \left( p\!\!\!/ - m \right) | q + \rangle,
 & &
 \bar{v}(-) = \frac{1}{\langle q + | p^\flat - \rangle} \langle q + | \left( p\!\!\!/ - m \right).
\eq
We label the helicities as if all particles were outgoing. 
As a consequence, the spinors $u(\lambda)$ and $\bar{v}(\lambda)$, which correspond
to particles with incoming momentum, have a reversed helicity assignment.
In the massless limit the definition reduces to
\bq
 u(-) = v(-) = | p+ \rangle,
 &&
 \bar{u}(+) = \bar{v}(+) = \langle p+ |, 
 \nonumber \\
 u(+) = v(+) = | p- \rangle,
 & &
 \bar{u}(-) = \bar{v}(-) = \langle p- |,
\eq
and the spinors are independent of the reference spinors $|q+\rangle$ and $\langle q+|$.

For massive fermions the reference spinors are related to the quantisation axis of the spin for 
this fermion, and the individual amplitudes with label $+$ or $-$ will therefore depend on the
reference spinors $|q+\rangle$ and $\langle q+|$.
It should be noted that the spinors for the massive fermions depend both on $|q+\rangle$ and $\langle q+|$:
For the spinors with helicity ``$+$'' there is an explicit dependence on $|q+\rangle$, while the dependence on $\langle q+|$
enters implicitly through $p^\flat$.
In a similar way we find that the spinors with helicity ``$-$'' have an explicit dependence on $\langle q+|$, 
while the dependence on $|q+\rangle$ enters implicitly through $p^\flat$.

It is easy to relate helicity amplitudes of massive quarks
corresponding to one choice of reference spinors to another set of reference spinors.
If $|\tilde{q}+\rangle$ and $\langle \tilde{q}+|$ is a second pair of reference spinors
we have the following transformation law 
\bq
\label{rotate_spin_1}
\left( \begin{array}{c}
 \bar{u}(+,\tilde{q}) \\
 \bar{u}(-,\tilde{q}) \\
 \end{array} \right)
 & = &
 \left( \begin{array}{cc}
 c_{11} & c_{12} \\
 c_{21} & c_{22} \\
 \end{array} \right)
\left( \begin{array}{c}
 \bar{u}(+,q) \\
 \bar{u}(-,q) \\
 \end{array} \right),
\eq
where
\bq
 c_{11} = \frac{\langle \tilde{q}- | p\!\!\!/ | q- \rangle}{\langle \tilde{q} \tilde{p}^\flat \rangle [ p^\flat q ]},
 \;\;\;
 c_{12} = \frac{m \langle \tilde{q} q \rangle}{\langle \tilde{q} \tilde{p}^\flat \rangle \langle p^\flat q \rangle},
 \;\;\;
 c_{21} = \frac{m [ \tilde{q} q ]}{[ \tilde{q} \tilde{p}^\flat] [ p^\flat q ]},
 \;\;\;
 c_{22} = \frac{\langle \tilde{q}+ | p\!\!\!/ | q+ \rangle}{[ \tilde{q} \tilde{p}^\flat] \langle p^\flat q \rangle}.
\eq
Here, $\tilde{p}^\flat$ denotes the projection onto a light-like four-vector with respect to the reference vector
$\frac{1}{2} \langle \tilde{q}+ | \gamma^\mu | \tilde{q} + \rangle$.
Similar, we have for an amplitude with an incoming massive quark
\bq
\label{rotate_spin_2}
\left( \begin{array}{c}
 u(+,\tilde{q}) \\
 u(-,\tilde{q}) \\
 \end{array} \right)
 & = &
 \left( \begin{array}{cc}
 c_{11} & -c_{12} \\
 -c_{21} & c_{22} \\
 \end{array} \right)
\left( \begin{array}{c}
 u(+,q) \\
 u(-,q) \\
 \end{array} \right).
\eq
Similar formulae exist for the spinors $v$ and $\bar{v}$ and can be obtained
by the substitution $u \rightarrow v$, $\bar{u} \rightarrow \bar{v}$ and $m \rightarrow -m$.

For the polarisation vectors of a gluon with momentum $p$  we take
\bq
\eps_{\mu}^{+} = \frac{\langle p+|\gamma_{\mu}|q+\rangle}{\sqrt{2} \langle q- | p + \rangle},
 & &
\eps_{\mu}^{-} = \frac{\langle q+|\gamma_{\mu}|p+\rangle}{\sqrt{2} \langle p + | q - \rangle}.
\eq
The dependence on the reference spinors $|q+\rangle$ and $\langle q + |$, which enters through the gluon polarisation vectors
will drop out in gauge invariant quantities.

\subsection{Decomposition of massive four-vectors into light-like four-vectors}
\label{sect:definition_l1_l2}

The external momenta of particles $1$ and $n$ may be massive or massless.
In the case where they are massive (either one of them or both) we would like to write them as a linear
combination of two light-like four-vectors $l_1$ and $l_n$.
The two light-like four-vectors $l_1$ and $l_n$ are constructed as follows \cite{delAguila:2004nf,vanHameren:2005ed}:
If $p_1$ and $p_n$ are massless, $l_1$ and $l_n$ are given by
\bq
 l_1 = p_1, 
 & &
 l_n = p_n.
\eq
If $p_1$ is massless, but $p_n$ is massive one has
\bq
 l_1 = p_1, 
 \;\;\;
 l_n = -\alpha_1 p_1 + p_n,
 \;\;\;
 \alpha_1 = \frac{p_n^2}{2p_1 p_n}.
\eq
The inverse formula is given by
\bq
 p_1 = l_1, 
 & &
 p_n = \alpha_1 l_1 + l_n.
\eq
If both $p_1$ and $p_n$ are massive, one has
\bq
 l_1 = \frac{1}{1-\alpha_1 \alpha_n} \left( p_1 - \alpha_n p_n \right), 
 & &
 l_n = \frac{1}{1-\alpha_1 \alpha_n} \left( -\alpha_1 p_1 + p_n \right).
\eq
$\alpha_1$ and $\alpha_n$ are given by
\bq
\label{def_alpha}
 \alpha_1 = \frac{2p_1p_n-\mathrm{sign}(2p_1p_n)\sqrt{\Delta}}{2p_1^2},
 & &
 \alpha_n = \frac{2p_1p_n-\mathrm{sign}(2p_1p_n)\sqrt{\Delta}}{2p_n^2}.
\eq
Here,
\bq
\Delta & = & \left( 2p_1p_n \right)^2 - 4p_1^2 p_n^2.
\eq
The signs are chosen in such away that the massless limit $p_1^2 \rightarrow 0$ (or $p_n^2 \rightarrow 0$)
is approached smoothly.
The inverse formulae are given by
\bq
\label{decompmom}
 p_1 = l_1 + \alpha_n l_n, 
 & &
 p_n = \alpha_1 l_1 + l_n.
\eq
The light-like four-vectors $l_1$ and $l_n$ define massless spinors
$|l_1+ \rangle$, $\langle l_1+ |$, $|l_n+ \rangle$ and $\langle l_n+ |$.

\subsection{On the choice of the reference spinors}

Particles $1$ and $n$ may be massive quarks or anti-quarks and we have to make a choice for the reference spinors.
In the massless case, the amplitude will be independent of the choice of the reference spinors 
and this section is of no further relevance.
However, if particle $1$ (or particle $n$) is massive,
the choice of the reference spinors will define the spin quantisation axis and the amplitude will depend on this choice.
It is always possible to convert to a different basis with the help of eqs.~(\ref{rotate_spin_1})-(\ref{rotate_spin_2}).

In section~(\ref{sect:definition_l1_l2}) we have constructed the spinors 
$|l_1+ \rangle$, $|l_n+ \rangle$, $\langle l_1+ |$ and $\langle l_n+ |$.
For generic momenta $p_1$ and $p_n$, the ket-spinors $|l_1+ \rangle$ and $|l_n+ \rangle$
span the two-dimensional space of holomorphic spinors (or ket-plus-spinors).
Similarly, the bra-spinors $\langle l_1+ |$ and $\langle l_n+ |$ span the space of anti-holomorphic spinors (or bra-plus-spinors).
For particle $1$ we parametrise the reference spinors $|q_1+ \rangle$ and $\langle q_1+ |$ as
\bq
 \left| q_1 + \right\rangle \;\; = \;\; \left| l_n + \right\rangle + \lambda_1 \left| l_1 + \right\rangle,
 & &
 \left\langle q_1 + \right| \;\; = \;\; \left\langle l_n + \right| + \lambda_1 \left\langle l_1 + \right|,
\eq
with one parameter $\lambda_1$.
For particle $n$ we parametrise the reference spinors $|q_n+ \rangle$ and $\langle q_n+ |$ as
\bq
 \left| q_n + \right\rangle \;\; = \;\; \left| l_1 + \right\rangle + \lambda_n \left| l_n + \right\rangle,
 & &
 \left\langle q_n + \right| \;\; = \;\; \left\langle l_1 + \right| + \lambda_n \left\langle l_n + \right|,
\eq
with one further parameter $\lambda_n$.
Working out $|p_1^\flat+\rangle$, $|p_n^\flat+\rangle$, $\langle p_1^\flat+ |$ and $\langle p_n^\flat+|$ one finds
\bq
 \left| p_1^\flat + \right\rangle
 \;\; = \;\;
 c_1 \left( \left| l_1 + \right\rangle - \alpha_n \lambda_1 \left| l_n + \right\rangle \right),
 & &
 \left\langle p_1^\flat + \right|
 \;\; = \;\;
 c_1 \left( \left\langle l_1 + \right| - \alpha_n \lambda_1 \left\langle l_n + \right| \right),
 \nonumber \\
 \left| p_n^\flat + \right\rangle
 \;\; = \;\;
 c_n \left( \left| l_n + \right\rangle - \alpha_1 \lambda_n \left| l_1 + \right\rangle \right),
 & &
 \left\langle p_n^\flat + \right|
 \;\; = \;\;
 c_n \left( \left\langle l_n + \right| - \alpha_1 \lambda_n \left\langle l_1 + \right| \right),
\eq
with
\bq
\label{def_c}
 c_1 
 \;\; = \;\;
 \frac{1}{\sqrt{1+\alpha_n\lambda_1^2}},
 & &
 c_n 
 \;\; = \;\;
 \frac{1}{\sqrt{1+\alpha_1\lambda_n^2}}.
\eq
The variables $\alpha_1$ and $\alpha_n$ have been defined in eq.~(\ref{def_alpha}).
We will use the freedom to choose $\lambda_1$ and $\lambda_n$
to compensate a restriction on the possible BCFW-shifts for massive particles.
The shifted spinors of the massive particles have to satisfy the Dirac equation, as well as orthogonality and completeness relations.
This restricts the $z$-dependent part to be proportional to the corresponding reference spinors \cite{Schwinn:2007ee}.
This means if we shift a massive spinor
\bq
 \hat{u}(-) & = & u(-) - z \left| \eta + \right\rangle,
\eq
the spinor $| \eta + \rangle$ has to be proportional to $| q + \rangle$:
\bq
 \left| \eta + \right\rangle & = &  \kappa \left| q + \right\rangle.
\eq
Similarly, if we shift
\bq
 \hat{\bar{u}}(+) & = & \bar{u}(+) + z \left\langle \eta + \right|,
\eq
we have to require that
\bq
 \left\langle \eta + \right| & = & \kappa \left\langle q + \right|.
\eq

\subsection{BCFW-shifts}

We now provide the explicit formulae for the three-particle shifts.
We have to consider all possible helicity configurations for the three particles $1$, $2_g$ and $n$.
In all cases the deformations are defined in such a way, that the external polarisations spinors and vectors give the best possible large $z$-behaviour.
This is a $z^{-3}$-behaviour if the three particles $1$, $2_g$ and $n$ are all gluons,
a $z^{-2}$-behaviour if one of them is a quark or an anti-quark and
a $z^{-1}$-behaviour if two of them are quarks or anti-quarks.
As particle $2_g$ is required to be a gluon, the case where all three particles are quarks or anti-quarks is not possible.
\begin{table}
\begin{center}
\begin{tabular}{|l|l|}
\hline
 particles $1$, $2_g$ and $n$ & large $z$-behaviour \\
 \hline
 $3$ gluons & $z^{-3}$ \\
 $2$ gluons, $1$ quark/anti-quark & $z^{-2}$ \\
 $1$ gluon, $2$ quarks/anti-quarks & $z^{-1}$ \\
 \hline
\end{tabular}
\caption{\label{table1}
The large $z$-behaviour of the external polarisations under the three-particle shifts.
}
\end{center}
\end{table}
The large $z$-behaviour of the external polarisations is summarised in table~(\ref{table1}).

\subsubsection{The helicity configuration $1^+, 2_g^+, n^-$}
\label{sect:case1}

For the helicity configuration $1^+, 2_g^+, n^-$ 
we shift $u_1(-)$, $u_2(-)$ and $\bar{u}_n(+)$, while $\bar{u}_1(+)$, $\bar{u}_2(+)$ and $u_n(-)$ remain unchanged: 
\bq
\label{shift_++-}
 \hat{u}_1(-) = u_1(-) - z y_1 | p_n^\flat + \rangle,
 & &
 \hat{\bar{u}}_n(+) = \bar{u}_n(+) + z y_1 \langle p_1^\flat + | + z y_2 \langle p_2 + |,
 \nonumber \\
 \hat{u}_2(-) = u_2(-) - z y_2 | p_n^\flat + \rangle,
 & &
\eq
where $y_1$ and $y_2$ are two non-zero constants.
For massive particles we have to require, that the shift is proportional to the corresponding reference
spinors.
Therefore we have to require that the system of equations
\bq
 \kappa_1 \left| q_1 + \right\rangle 
 \;\; = \;\;
 y_1 \left| p_n^\flat + \right\rangle,
 & &
 \kappa_n \left\langle q_n + \right| 
 \;\; = \;\; 
 y_1 \left\langle p_1^\flat + \right| + y_2 \left\langle p_2 + \right|,
\eq
has a solution for some non-zero constants $\kappa_1$ and $\kappa_2$. 
In appendix~(\ref{appendix_++-}) we show that this is the case.
The spinors $\hat{u}_1(-)$ and $\hat{\bar{u}}_1(+)$ correspond to an on-shell particle with mass $m_1$ and four-momentum
\bq
 \hat{p}_1^\mu & = & p_1^\mu - \frac{1}{2} z y_1 \left\langle p_1^\flat+ \left| \gamma^\mu \right| p_n^\flat+ \right\rangle.
\eq
The spinors $\hat{u}_2(-)$ and $\hat{\bar{u}}_2(+)$ correspond to an on-shell particle with zero mass and four-momentum
\bq
 \hat{p}_2^\mu & = & p_2^\mu - \frac{1}{2} z y_2 \left\langle p_2+ \left| \gamma^\mu \right| p_n^\flat+ \right\rangle.
\eq
The spinors $\hat{u}_n(-)$ and $\hat{\bar{u}}_n(+)$ correspond to an on-shell particle with mass $m_n$ and four-momentum
\bq
 \hat{p}_n^\mu & = & p_n^\mu + \frac{1}{2} z y_1 \left\langle p_1^\flat+ \left| \gamma^\mu \right| p_n^\flat+ \right\rangle
                             + \frac{1}{2} z y_2 \left\langle p_2+ \left| \gamma^\mu \right| p_n^\flat+ \right\rangle.
\eq

\subsubsection{The helicity configuration $1^+, 2_g^-, n^-$}
\label{sect:case2}

For the helicity configuration $1^+, 2_g^-, n^-$ 
we shift $u_1(-)$, $\bar{u}_2(+)$ and $\bar{u}_n(+)$, while $\bar{u}_1(+)$, $u_2(-)$ and $u_n(-)$ remain unchanged: 
\bq
\label{shift_+--}
 \hat{u}_1(-) = u_1(-) - z y_2 | p_2 + \rangle - z y_n | p_n^\flat + \rangle,
 & &
 \hat{\bar{u}}_2(+) = \bar{u}_2(+) + z y_2 \langle p_1^\flat + |,
 \nonumber \\
 & &
 \hat{\bar{u}}_n(+) = \bar{u}_n(+) + z y_n \langle p_1^\flat + |,
\eq
where $y_2$ and $y_n$ are two non-zero constants.
For massive particles we have to require that the system of equations
\bq
 \kappa_1 \left| q_1 + \right\rangle 
 \;\; = \;\;
 y_2 \left| p_2 + \right\rangle + y_n \left| p_n^\flat + \right\rangle,
 & &
 \kappa_n \left\langle q_n + \right| 
 \;\; = \;\; 
 y_n \left\langle p_1^\flat + \right|,
\eq
has a solution for some non-zero constants $\kappa_1$ and $\kappa_n$. 
In appendix~(\ref{appendix_+--}) we show that this is the case.
The spinors $\hat{u}_1(-)$ and $\hat{\bar{u}}_1(+)$ correspond to an on-shell particle with mass $m_1$ and four-momentum
\bq
 \hat{p}_1^\mu & = & p_1^\mu - \frac{1}{2} z y_2 \left\langle p_1^\flat+ \left| \gamma^\mu \right| p_2+ \right\rangle
                             - \frac{1}{2} z y_n \left\langle p_1^\flat+ \left| \gamma^\mu \right| p_n^\flat+ \right\rangle.
\eq
The spinors $\hat{u}_2(-)$ and $\hat{\bar{u}}_2(+)$ correspond to an on-shell particle with zero mass and four-momentum
\bq
 \hat{p}_2^\mu & = & p_2^\mu + \frac{1}{2} z y_2 \left\langle p_1^\flat+ \left| \gamma^\mu \right| p_2+ \right\rangle.
\eq
The spinors $\hat{u}_n(-)$ and $\hat{\bar{u}}_n(+)$ correspond to an on-shell particle with mass $m_n$ and four-momentum
\bq
 \hat{p}_n^\mu & = & p_n^\mu + \frac{1}{2} z y_n \left\langle p_1^\flat+ \left| \gamma^\mu \right| p_n^\flat+ \right\rangle.
\eq

\subsubsection{The helicity configuration $1^+, 2_g^-, n^+$}
\label{sect:case3}

For the helicity configuration $1^+, 2_g^-, n^+$ 
we shift $u_1(-)$, $\bar{u}_2(+)$ and $u_n(-)$, while $\bar{u}_1(+)$, $u_2(-)$ and $\bar{u}_n(+)$ remain unchanged: 
\bq
\label{shift_+-+}
 \hat{u}_1(-) = u_1(-) - z y_1 | p_2 + \rangle,
 & &
 \hat{\bar{u}}_2(+) = \bar{u}_2(+) + z y_1 \langle p_1^\flat + | + z y_n \langle p_n^\flat + |,
 \nonumber \\
 \hat{u}_n(-) = u_n(-) - z y_n | p_2 + \rangle,
 & &
\eq
where $y_1$ and $y_n$ are two non-zero constants.
For massive particles we choose
\bq
 \left| q_1+ \right\rangle 
 \;\; = \;\;
 \left| q_n+ \right\rangle 
 \;\; = \;\;
 \left| p_2+ \right\rangle,
 & &
 \left\langle q_1+ \right|
 \;\; = \;\;
 \left\langle q_n+ \right|
 \;\; = \;\;
 \left\langle p_2+ \right|
\eq
as reference spinors.
The spinors $\hat{u}_1(-)$ and $\hat{\bar{u}}_1(+)$ correspond to an on-shell particle with mass $m_1$ and four-momentum
\bq
 \hat{p}_1^\mu & = & p_1^\mu - \frac{1}{2} z y_1 \left\langle p_1^\flat+ \left| \gamma^\mu \right| p_2+ \right\rangle.
\eq
The spinors $\hat{u}_2(-)$ and $\hat{\bar{u}}_2(+)$ correspond to an on-shell particle with zero mass and four-momentum
\bq
 \hat{p}_2^\mu & = & p_2^\mu + \frac{1}{2} z y_1 \left\langle p_1^\flat+ \left| \gamma^\mu \right| p_2+ \right\rangle
                             + \frac{1}{2} z y_n \left\langle p_n^\flat+ \left| \gamma^\mu \right| p_2+ \right\rangle.
\eq
The spinors $\hat{u}_n(-)$ and $\hat{\bar{u}}_n(+)$ correspond to an on-shell particle with mass $m_n$ and four-momentum
\bq
 \hat{p}_n^\mu & = & p_n^\mu - \frac{1}{2} z y_n \left\langle p_n^\flat+ \left| \gamma^\mu \right| p_2+ \right\rangle.
\eq

\subsubsection{The helicity configuration $1^+, 2_g^+, n^+$}
\label{sect:case4}

For the helicity configuration $1^+, 2_g^+, n^+$ 
we shift $u_1(-)$, $u_2(-)$ and $u_n(-)$, while $\bar{u}_1(+)$, $\bar{u}_2(+)$ and $\bar{u}_n(+)$ remain unchanged: 
\bq
\label{shift_+++}
 \hat{u}_1(-) & = & u_1(-) - z \left[ p_2 p_n^\flat \right] \; | \eta + \rangle,
 \nonumber \\
 \hat{u}_2(-) & = & u_2(-) - z \left[ p_n^\flat p_1^\flat \right] \; | \eta + \rangle,
 \nonumber \\
 \hat{u}_n(-) & = & u_n(-) - z \left[ p_1^\flat p_2 \right] \; | \eta + \rangle.
\eq
Here, $| \eta + \rangle$ is an arbitrary spinor.
For massive particles we choose
\bq
 \left| q_1+ \right\rangle 
 \;\; = \;\;
 \left| q_n+ \right\rangle 
 \;\; = \;\;
 \left| \eta+ \right\rangle,
 & &
 \left\langle q_1+ \right|
 \;\; = \;\;
 \left\langle q_n+ \right|
 \;\; = \;\;
 \left\langle \eta+ \right|
\eq
as reference spinors.
The spinors $\hat{u}_1(-)$ and $\hat{\bar{u}}_1(+)$ correspond to an on-shell particle with mass $m_1$ and four-momentum
\bq
 \hat{p}_1^\mu & = & p_1^\mu - \frac{1}{2} z \left[ p_2 p_n^\flat \right] \left\langle p_1^\flat+ \left| \gamma^\mu \right| \eta+ \right\rangle.
\eq
The spinors $\hat{u}_2(-)$ and $\hat{\bar{u}}_2(+)$ correspond to an on-shell particle with zero mass and four-momentum
\bq
 \hat{p}_2^\mu & = & p_2^\mu - \frac{1}{2} z \left[ p_n^\flat p_1^\flat \right] \left\langle p_2+ \left| \gamma^\mu \right| \eta+ \right\rangle.
\eq
The spinors $\hat{u}_n(-)$ and $\hat{\bar{u}}_n(+)$ correspond to an on-shell particle with mass $m_n$ and four-momentum
\bq
 \hat{p}_n^\mu & = & p_n^\mu - \frac{1}{2} z \left[ p_1^\flat p_2 \right] \left\langle p_n^\flat+ \left| \gamma^\mu \right| \eta+ \right\rangle.
\eq
Momentum conservation is satisfied due to the Schouten identity.

\subsubsection{The remaining helicity configurations}

The shifts for the helicity configurations 
\bq
 (1^-, 2_g^-, n^+), (1^-, 2_g^+, n^+), (1^-, 2_g^+, n^-), (1^-, 2_g^-, n^-)
\eq
can be obtained from the helicity configurations
\bq
 (1^+, 2_g^+, n^-), (1^+, 2_g^-, n^-), (1^+, 2_g^-, n^+), (1^+, 2_g^+, n^+)
\eq
by exchanging holomorphic and anti-holomorphic spinors.
 
% ----------------------------------------------------------------------------------
\section{Large-$z$ behaviour}
\label{sect:large_z_behaviour}

We consider $I_n(z)$ for $n \ge 4$.
$I_n(z)$ is a rational function in $z$. 
We have to show that $I_n(z)$ falls off at $z=\infty$ at least with $1/z$.
We will distinguish the cases, where the three particles $1$, $2_g$ and $n$ are 
\begin{description}
\item{(i)} three gluons,
\item{(ii)} two gluons and one quark/anti-quark,
\item{(iii)} one gluon and two quarks/anti-quarks, not belonging to the same fermion line or
\item{(iv)} one gluon and a quark-anti-quark-pair belonging to the same fermion line.
\end{description}
Let us recall the definition of $I_n(z)$:
\bq
 I_n\left(z\right)
 & = &
 \sum\limits_{i=2}^{n-1}
  \left( \sum\limits_{j=i+1}^n 2 \hat{p}_2 \hat{p}_j \right)
  A_n\left(\hat{1},3,...,i,\hat{2}_g,i+1,...,n-1,\hat{n}\right).
\eq
We note that the factors $(2 \hat{p}_2 \hat{p}_j)$ are at the worst linear in $z$.
A sufficient condition is to show that each amplitude $A_n(\hat{1},3,...,i,\hat{2}_g,i+1,...,n-1,\hat{n})$
falls off at $z=\infty$ at least with $1/z^2$.
We will show that this holds for the cases (i)-(iii).

However, a $1/z^2$-fall-off behaviour of the amplitudes is not a necessary condition.
In fact, in the case (iv) the amplitudes fall off only with $1/z$. 
In this case we show through a more sophisticated argument that the full sum $I_n(z)$ falls off at $z=\infty$ with $1/z$.

\subsection{Three gluons}

Let us start with the case $(1_g,2_g,n_g)$. 
The external polarisation vectors contribute
a factor $z^{-3}$.
The most critical contribution from the vertices and propagators comes from diagrams, where there are
only three-gluon vertices along the $z$-flow.
For these diagrams there will be along the $z$-flow always one more three-gluon vertex as there are propagators, giving 
a net factor of $z^{1}$.
Therefore we obtain from these diagrams a total contribution of $z^{-3} \cdot z = z^{-2}$.
If internally a gluon propagator is replaced by a quark propagator, we have to change at least
two three-gluon vertices into quark-gluon vertices. This improves the estimate by a factor $1/z$.
Similarly, the replacement of one three-gluon vertex by a four-gluon vertex results in an improvement in the $z$-behaviour
by a factor $1/z$.
We therefore conclude, that the amplitude falls off at $z=\infty$ at least with $1/z^2$.

\subsection{Two gluons and one quark/anti-quark}

The arguments for the cases $(1_{q/\bar{q}},2_g,n_g)$ and $(1_g,2_g,n_{q/\bar{q}})$
are very similar to the three gluon case.
Although the external polarisations contribute now only a factor $z^{-2}$, the estimate
from the vertices and the propagators is now $z^0$.
Again, the worst diagrams are the ones with a maximal number of three-gluon vertices along the $z$-flow.
However, in the case at hand we must have at least one quark-gluon-vertex along the $z$-flow, improving
the estimate by a factor $1/z$.
Again we see that the amplitude falls off at $z=\infty$ at least with $1/z^2$.

\subsection{One gluon and two quarks/anti-quarks, not belonging to the same fermion line}

Let us now discuss the case $(1_{q/\bar{q}},2_g,n_{q'/\bar{q}'})$ 
with one gluon and two quarks/anti-quarks, where the two fermions do not belong to the same fermion line.
This sub-case is straightforward: 
Although the external polarisations contribute now only a factor $z^{-1}$, the estimate
from the vertices and the propagators is now $z^{-1}$.
This is due to the fact that we have to change yet another three-gluon vertex into a quark-gluon vertex.
Again one concludes that the amplitude falls off at $z=\infty$ at least with $1/z^2$.

\subsection{One gluon and a quark-anti-quark-pair belonging to the same fermion line}

The case $(1_q,2_g,n_{\bar{q}})$ and $(1_{\bar{q}},2_g,n_q)$, where the two fermions belong to the same fermion line,
are more complicated.
Power-counting gives now a factor $z^{-1}$ from the external polarisations and
a factor $z^0$ from the vertices and propagators.
The individual amplitudes fall off as $1/z$ for large $z$.
In this case we show, that the sum $I_n(z)$ falls off as $1/z$ for large $z$.
The worst diagrams are the ones, where the $z$-flow of gluon $2_g$ goes only through three-gluon vertices
before it couples to the quark line.
We have to show that in the sum the leading $z$-behaviour of these diagrams actually vanishes.
For the leading $z$-behaviour we can use an argument of Arkani-Hamed and Kaplan \cite{ArkaniHamed:2008yf}:
For large $z$ we may view particles $1$, $2_g$ and $n$ as highly energetic particles moving in a soft background.
All vertices along the $z$-flow reduce in this limit to eikonal factors, except the one where the
three branches of the $z$-flow meet.
In order to see this let us start from particle $2$ and consider the first vertex particle $2$ meets.
This three-gluon vertex couples particle $2$, a current containing only soft particles
\bq
 J_\mu^{\mathrm{soft}}
 & = &
 J_\mu^{\mathrm{soft}}\left(k+1,...,l\right)
\eq
and a current containing the other hard particles $1$ and $n$
\bq
 \hat{J}_\mu^{\mathrm{hard}}
 & = &
 \hat{J}_\mu^{\mathrm{hard}}\left(\hat{1},3,...,k,l+1,...,\hat{n}\right).
\eq
In the Feynman rule for the three-gluon vertex we only have to keep the $z$-dependent terms, 
yielding for the cyclic order $2_g, J^{\mathrm{soft}}, \hat{J}^{\mathrm{hard}}$
\bq
 i \left[ 
  - \left( \hat{\eps}_2 \cdot J^{\mathrm{soft}} \right) \left( \hat{p}_{\mathrm{hard}} \cdot \hat{J}^{\mathrm{hard}} \right)
  + \left( \hat{p}_2 \cdot \hat{\eps}_2 \right) \left( J^{\mathrm{soft}} \cdot \hat{J}^{\mathrm{hard}} \right)
  - 2 \left( \hat{\eps}_2 \cdot \hat{J}^{\mathrm{hard}} \right) \left( \hat{p}_2 \cdot J^{\mathrm{soft}} \right)
 \right].
\eq
The contraction of $\hat{p}_2$ with $\hat{\eps}_2$ vanishes: $\hat{p}_2 \cdot \hat{\eps}_2=0$.
Furthermore, the current $\hat{J}^{\mathrm{hard}}$ is conserved and we have $\hat{p}_{\mathrm{hard}} \cdot \hat{J}^{\mathrm{hard}} = 0$.
This leaves the eikonal contribution
\bq
  \left( \hat{\eps}_2 \cdot \hat{J}^{\mathrm{hard,amputated}} \right)
  \left( - \frac{2 \hat{p}_2 \cdot J^{\mathrm{soft}}}{\left(\hat{p}_2 +p_{\mathrm{soft}}\right)^2 } \right),
\eq
with
\bq
 \hat{J}^{\mathrm{hard,amputated}}
 & = &
 i
 \left(\hat{p}_2 +p_{\mathrm{soft}}\right)^2
 \hat{J}^{\mathrm{hard}}.
\eq
We may then repeat the argument with the next three-gluon vertex.
A similar argument can be given for the $z$-flow along the quark line.
Let us start at particle $1$ and let us assume that this particle is a quark.
We consider the first vertex particle $1$ meets. This is a quark-gluon vertex,
connecting particle $1$, 
a gluon current containing only soft particles
\bq
 J_\mu^{\mathrm{soft}}
 & = &
 J_\mu^{\mathrm{soft}}\left(3,...,k\right)
\eq
and a hard spinorial current containing the other two hard particles $2$ and $n$:
\bq
 \hat{V}^{\mathrm{hard}}
 & = &
 \hat{V}^{\mathrm{hard}}\left(k+1,...,\hat{2},...,n-1,\hat{n}\right).
\eq
Let us further define the hard amputated spinorial current as
\bq
 \hat{V}^{\mathrm{hard}}
 & = &
 i \frac{\left(\hat{p}\!\!\!/_1+p\!\!\!/_{\mathrm{soft}}\right) + m}{\left(\hat{p}\!\!\!/_1+p\!\!\!/_{\mathrm{soft}}\right)^2-m^2}
  \hat{V}^{\mathrm{hard,amputated}}.
\eq
Again we may neglect soft momenta in the numerator and we find
\bq
\lefteqn{
 -
 \hat{\bar{u}}_1 \gamma^\mu
 \frac{\hat{p}\!\!\!/_1 + m}{\left(\hat{p}\!\!\!/_1+p\!\!\!/_{\mathrm{soft}}\right)^2-m^2}
  \hat{V}^{\mathrm{hard,amputated}}
 J_\mu^{\mathrm{soft}}
 = }
 & &
 \nonumber \\
 & &
 \left( - \frac{2 \hat{p}_1 \cdot J^{\mathrm{soft}}}{\left(\hat{p}\!\!\!/_1+p\!\!\!/_{\mathrm{soft}}\right)^2-m^2} \right)
 \hat{\bar{u}}_1 
  \hat{V}^{\mathrm{hard,amputated}}
 + 
 \frac{\hat{\bar{u}}_1 \left( \hat{p}\!\!\!/_1 - m \right) \gamma^\mu \hat{V}^{\mathrm{hard,amputated}}}{\left(\hat{p}\!\!\!/_1+p\!\!\!/_{\mathrm{soft}}\right)^2-m^2}
 J_\mu^{\mathrm{soft}}.
\eq
In the first term on the right-hand side we recognise an eikonal factor, the second term vanishes due to the Dirac equation.
As before, we may repeat the argument with the next quark-gluon vertex.

The argument for the branch with the external anti-quark at position $n$ is identical and not repeated here.
The $(1_{\bar{q}},2_g,n_q)$-case is very similar and not discussed in detail.

The eikonal factors go to a constant for large $z$ and we are left with a quark-gluon vertex
contracted for the $(1_q,2_g,n_{\bar{q}})$-case with $\hat{\bar{u}}_1$, $\hat{\eps}_2$ and $\hat{v}_n$.
Let us denote this contribution by
\bq
\label{O_3}
 O_3 
 & = &
 i \hat{\bar{u}}_1 \gamma_\mu \hat{v}_n \hat{\eps}_2^\mu.
\eq
The quantity $O_3$ falls off like $1/z$ for large $z$. 
It is important to note, that $O_3$ occurs in every amplitude contributing to $I_n(z)$ in the 
$(1_q,2_g,n_{\bar{q}})$-case.
It may therefore be taken out of the sum, and we have to show that the remaining sum goes to a constant for large $z$.
The remaining sum involves only the Lorentz invariants $2\hat{p}_2\hat{p}_j$ and the eikonal factors.
The proof is given in appendix~(\ref{appendix:eikonal}).

% ----------------------------------------------------------------------------------
\section{The proof by induction}
\label{sect:induction}

In this section we prove the fundamental BCJ-relation by induction.
With the preparations of section~(\ref{sect:deformation}) and section~(\ref{sect:large_z_behaviour}) 
we can do this independently of the helicity configurations
and the masses.
This is possible, since we have for $I_n(z)$ 
for all helicity configurations and all masses a $1/z$-behaviour for large $z$.
However, we would like to point out one subtle point for massive quarks:
We would like to show that the fundamental BCJ-relations holds for all helicities of the massive quark.
The naive way to show this would be to fix a spin quantisation axis through a choice of reference spinors
$|q+\rangle$ and $\langle q+|$ and to show the BCJ-relation for the helicities ``$+$'' and ``$-$'' with respect
to these reference spinors.
This is not what we are doing. 
The attentive reader of sections~(\ref{sect:case1})-(\ref{sect:case4})
might have noticed, that the ``$+$''- and ``$-$''-helicities refer to different reference spinors.
This is o.k., since amplitudes with different spin quantisation axes are related through 
eq.~(\ref{rotate_spin_1}) and eq.~(\ref{rotate_spin_2}).
Therefore it is sufficient to know two independent amplitudes 
(say ``$+$''-helicity with respect to $q$ and ``$-$''-helicity with respect to $\tilde{q}$)
in order to know all amplitudes with spin quantisation axes $q$ and $\tilde{q}$.
This remark applies to each external particle individually and covers all possible cases for
the external particles $1$ and $n$, where we can have out of these two particles either zero, one or two
massive particles. In the latter case the masses may be equal or not.

\subsection{Induction start: The case $n=3$}

To start the proof by induction we consider the case $n=3$.
Throughout this paper we work with complex external momenta.
The external momenta satisfy momentum conservation
\bq
\label{momentum_conservation_3}
 p_1 + p_2 + p_3 & = & 0,
\eq
and the on-shell conditions
\bq
\label{on_shell_3}
 p_1^2 = m^2,
 \;\;\;\;\;\;
 p_2^2 = 0,
 \;\;\;\;\;\;
 p_3^2 = m^2.
\eq
Particle $2_g$ will always be a gluon and is therefore massless.
Particles $1$ and $3$ may be massless or massive. In the massive case, particles $1$ and $3$ are necessarily a quark-anti-quark pair of the same
flavour.
Therefore particles $1$ and $3$ will have the same mass $m$.
For $n=3$ external particles the  momentum configurations satisfying eq.~(\ref{momentum_conservation_3}) and eq.~(\ref{on_shell_3}) are in general complex.
The fundamental BCJ-relation reduces to
\bq
  2 p_2 p_3 \; A_3\left(1,2_g,3\right)
 & = & 0.
\eq
For generic external momenta $A_3(1,2_g,3)$ is finite and
\bq
 2 p_2 p_3 
 \;\; = \;\;
 \left(p_2+p_3\right)^2 - m^2
 \;\; = \;\;
 p_1^2 - m^2
 \;\; = \;\;
 0.
\eq

\subsection{The induction step}

We now show that
\bq
 I_j\left(0\right) & = & 0
\eq
holds for $j=n$, provided it holds for all $j < n$.
We start from eq.~(\ref{eq_to_prove_2})
\bq
 I_n\left(0\right) 
 & = & 
 \frac{1}{2\pi i}
 \oint\limits_{z=0} \frac{dz}{z} I_n\left(z\right),
\eq
where the contour is a small counter-clockwise circle around $z=0$.
Deforming the contour to a large circle at infinity and the residues at the finite poles $z_\alpha \neq 0$ we
obtain
\bq
 I_n\left(0\right) 
 & = & 
 B
 - \sum\limits_{\alpha} \mathrm{res}\left(\frac{I_n\left(z\right)}{z}\right)_{z_\alpha},
\eq
where $B$ denotes the contribution from the large circle at infinity.
In section~(\ref{sect:large_z_behaviour}) we have shown that $I_n(z)$ falls off at least with
$1/z$ for z$\rightarrow \infty$ and therefore 
\bq
 B & = & 0.
\eq
It will be convenient to introduce the following notation for the various factorisation channels:
\bq
\lefteqn{
 A_n\left(\hat{1},2,...,k,\hat{P} | -\hat{P}, k+1,...,n-1,\hat{n}\right)
 =
} & & \nonumber \\
 & &
 \sum\limits_\lambda
 A_{k+1}\left(\hat{1},2,...,k,\hat{P}\right)
 \frac{i}{P^2}
 A_{n-k+1}\left(-\hat{P},k+1,...,n-1,\hat{n}\right),
\eq
together with the convention that the hatted quantities are evaluated at $z=z_\alpha$.
The sum is over the helicity of the intermediate particle.
Let us look at the $z$-momentum flow for a three-particle BCFW-shift.
For each diagram we may divide the $z$-dependent propagators into three segments.
Each segment starts at the common vertex, where the $z$-dependent momentum flow meets 
and goes outwards towards the particles 
$1$, $2_g$ and $n$.
We may use these segments to divide the finite residues into three groups and we write
\bq
 I_n\left(0\right)
 & = &
 R_1 + R_2 + R_n,
\eq
with
\bq
 R_1
 & = &
 \sum\limits_{i=2}^{n-1}
  \left( \sum\limits_{j=i+1}^n 2 \hat{p}_2 \hat{p}_j \right)
        \sum\limits_{k=3}^i A_n\left(\hat{1},3,...,k,\hat{P}|-\hat{P},k+1,...,i,\hat{2}_g,i+1,...,n-1,\hat{n}\right),
 \nonumber \\
 R_2
 & = &
 \sum\limits_{i=2}^{n-1}
  \left( \sum\limits_{j=i+1}^n 2 \hat{p}_2 \hat{p}_j \right)
 \nonumber \\
 & &
 \times
        \sum\limits_{k=2}^i 
        \sum\limits_{\substack{l=i \\ (k,l) \neq (i,i)}}^{n-1}
        A_n\left(k+1,...,i,\hat{2}_g,i+1,...,l,\hat{P}|-\hat{P},l+1,...,n-1,\hat{n},\hat{1},3,...,k\right),
 \nonumber \\
 R_n
 & = &
 \sum\limits_{i=2}^{n-1}
  \left( \sum\limits_{j=i+1}^n 2 \hat{p}_2 \hat{p}_j \right)
  \sum\limits_{k=i}^{n-2} A_n\left(\hat{1},3,...,i,\hat{2}_g,i+1,...,k,\hat{P}|-\hat{P},k+1,...,n-1,\hat{n}\right).
 \nonumber \\
\eq
Let us first look at $R_1$.
We may exchange the summation over $i$ and $k$ as
\bq
 \sum\limits_{i=2}^{n-1} \sum\limits_{k=3}^i f\left(i,k\right)
 & = &
 \sum\limits_{k=3}^{n-1} \sum\limits_{i=k}^{n-1} f\left(i,k\right).
\eq
One obtains
\bq
 R_1 & = &
 \sum\limits_{k=3}^{n-1} \sum\limits_{i=k}^{n-1}
  \left( \sum\limits_{j=i+1}^n 2 \hat{p}_2 \hat{p}_j \right)
   A_n\left(\hat{1},3,...,k,\hat{P}|-\hat{P},k+1,...,i,\hat{2}_g,i+1,...,n-1,\hat{n}\right).
 \;\;\;
\eq
We recognise the fundamental BCJ relation for $(n-k+2)$ external particles.
For $k \ge 3$ we have $(n-k+2)<n$.
We may therefore use the induction hypothesis and we conclude
\bq
 R_1 & = & 0.
\eq
The argument for $R_n$ is very similar. We first exchange the summation indices as
\bq
 \sum\limits_{i=2}^{n-1} \sum\limits_{k=i}^{n-2} f\left(i,k\right)
 & = &
 \sum\limits_{k=2}^{n-2} \sum\limits_{i=2}^{k} f\left(i,k\right).
\eq
We then obtain
\bq
 R_n
 & = &
 -
 \sum\limits_{k=2}^{n-2} \sum\limits_{i=2}^{k}
  \left( \sum\limits_{j=1}^{i} 2 \hat{p}_2 \hat{p}_j \right)
  A_n\left(\hat{1},3,...,i,\hat{2}_g,i+1,...,k,\hat{P}|-\hat{P},k+1,...,n-1,\hat{n}\right)
 \nonumber \\
 & = & 
 0.
\eq
Here we used momentum conservation in the sum over $j$. Again we recognise the fundamental BCJ relation in the form
of eq.~(\ref{fundamental_BCJ_relation_2}). It follows that $R_n$ vanishes.

Exchanging the summation indices for $R_2$ one obtains
\bq
 R_2
 & = &
 \sum\limits_{k=2}^{n-2}
 \sum\limits_{l=k+1}^{n-1}
 \\
 & & 
 \times
 \sum\limits_{i=k}^{l}
  \left( \sum\limits_{j=i+1}^n 2 \hat{p}_2 \hat{p}_j \right)
        A_n\left(k+1,...,i,\hat{2}_g,i+1,...,l,\hat{P}|-\hat{P},l+1,...,n-1,\hat{n},\hat{1},3,...,k\right).
 \nonumber
\eq
We may split the sum over $j$ as
\bq
 \sum\limits_{j=i+1}^n 2 \hat{p}_2 \hat{p}_j
 & = &
 \underbrace{\sum\limits_{j=i+1}^l 2 \hat{p}_2 \hat{p}_j}_{A}
 +
 \underbrace{\sum\limits_{j=l+1}^n 2 \hat{p}_2 \hat{p}_j}_{B}
\eq
The terms of type $A$ vanish again by the induction hypothesis
\bq
 \sum\limits_{i=k}^{l-1}
  \left( \sum\limits_{j=i+1}^l 2 \hat{p}_2 \hat{p}_j \right)
        A_{l-k+2}\left(\hat{P},k+1,...,i,\hat{2}_g,i+1,...,l\right)
 & = & 0.
\eq
Note that the sum over $i$ extends only to $(l-1)$, the case $i=l$ contributes only to the terms of type $B$.

For the terms of type $B$ the sum over $j$ is independent of $i$ and may be taken outside the sum over $i$.
The sum over $i$ vanishes then due to the $U(1)$-decoupling relation, given in eq.~(\ref{basic_Kleiss_Kuijf}):
\bq
  \left( \sum\limits_{j=l+1}^n 2 \hat{p}_2 \hat{p}_j \right)
 \sum\limits_{i=k}^{l}
        A_{l-k+2}\left(\hat{P},k+1,...,i,\hat{2}_g,i+1,...,l\right)
 & = & 0.
\eq
We therefore conclude that
\bq
 R_2
 & = & 0.
\eq
Putting the partial results for $R_1$, $R_2$ and $R_n$ together we find that
\bq
 I_n\left(0\right)
 & = &
 0.
\eq
This completes the proof.

% ----------------------------------------------------------------------------------
\section{Conclusions}
\label{sect:conclusions}

In this paper we provided a proof of the fundamental BCJ-relation, stated in eq.~(\ref{fundamental_BCJ_relation}),
for primitive tree amplitudes in QCD.
The proof holds for massless and massive quarks.
For the proof we used induction and BCFW-recursion relations.

% ----------------------------------------------------------------------------------

\begin{appendix}

\section{Reference spinors for massive particles}
\label{appendix:existence_solution}

\subsection{The helicity configuration $1^+, 2_g^+, n^-$}
\label{appendix_++-}

In this appendix we show that the system
\bq
 \kappa_1 \left| q_1 + \right\rangle 
 \;\; = \;\;
 y_1 \left| p_n^\flat + \right\rangle,
 & &
 \kappa_n \left\langle q_n + \right| 
 \;\; = \;\; 
 y_1 \left\langle p_1^\flat + \right| + y_2 \left\langle p_2 + \right|
\eq
has a solution. 
Expressing $\langle p_2+ |$ in terms of $\langle l_1+|$ and $\langle l_n+ |$
\bq
 \left\langle p_2 + \right|
 & = &
 \frac{\left[p_2 l_n \right]}{\left[ l_1 l_n \right]} \left\langle l_1 + \right|
 +
 \frac{\left[l_1 p_2 \right]}{\left[ l_1 l_n \right]} \left\langle l_n + \right|,
\eq
we obtain the system of equations
\bq
 \kappa_1 & = & y_1 c_n,
 \nonumber \\
 \kappa_1 \lambda_1 & = & - y_1 c_n \alpha_1 \lambda_n,
 \nonumber \\
 \kappa_n & = & y_1 c_1 + y_2 \frac{\left[p_2 l_n \right]}{\left[ l_1 l_n \right]},
 \nonumber \\
 \kappa_n \lambda_n & = & - y_1 c_1 \alpha_n \lambda_1 + y_2 \frac{\left[l_1 p_2 \right]}{\left[ l_1 l_n \right]}.
\eq
The variables $\alpha_1$ and $\alpha_2$ are defined in eq.~(\ref{def_alpha}),
the variables $c_1$ and $c_n$ are defined in eq.~(\ref{def_c}).
We look for a solution for the variables $\kappa_1$, $\kappa_n$, $y_1$, $y_2$, $\lambda_1$ and $\lambda_n$.
A possible solution is
\bq
 \kappa_1
 \;\; = \;\;
 \frac{c_n}{c_1},
 & &
 \kappa_n 
 \;\; = \;\;
 1 + \frac{2p_2 l_n}{2 l_1 l_n},
 \\
 y_1 
 \;\; = \;\;
 \frac{1}{c_1},
 & &
 y_2 
 \;\; = \;\;
 \frac{\left\langle p_2 l_n \right\rangle}{\left\langle l_1 l_n \right\rangle},
 \nonumber \\
 \lambda_1
 \;\; = \;\;
 \frac{p_n^2 \left\langle l_1 + \left| p\!\!\!/_2 \right| l_n + \right\rangle}
      {\left(2 l_1 l_n\right)^2 - p_1^2 p_n^2 + \left(2 l_1 l_n \right) \left(2 p_2 l_n\right)},
 & &
 \lambda_n
 \;\; = \;\;
 -
 \frac{2 l_1 l_n  \left\langle l_1 + \left| p\!\!\!/_2 \right| l_n + \right\rangle}
      {\left(2 l_1 l_n\right)^2 - p_1^2 p_n^2 + \left(2 l_1 l_n \right) \left(2 p_2 l_n\right)}.
 \nonumber
\eq
 
\subsection{The helicity configuration $1^+, 2_g^-, n^-$}
\label{appendix_+--}

In this appendix we show that the system
\bq
 \kappa_1 \left| q_1 + \right\rangle 
 \;\; = \;\;
 y_2 \left| p_2 + \right\rangle + y_n \left| p_n^\flat + \right\rangle,
 & &
 \kappa_n \left\langle q_n + \right| 
 \;\; = \;\; 
 y_n \left\langle p_1^\flat + \right|
\eq
has a solution. 
Expressing $| p_2+ \rangle$ in terms of $| l_1+\rangle$ and $| l_n+ \rangle$
\bq
 \left| p_2 + \right\rangle
 & = &
 \frac{\left\langle p_2 l_n \right\rangle}{\left\langle l_1 l_n \right\rangle} \left| l_1 + \right\rangle
 +
 \frac{\left\langle l_1 p_2 \right\rangle}{\left\langle l_1 l_n \right\rangle} \left| l_n + \right\rangle,
\eq
we obtain the system of equations
\bq
 \kappa_n & = & y_n c_1,
 \nonumber \\
 \kappa_n \lambda_n & = & - y_n c_1 \alpha_n \lambda_1,
 \nonumber \\
 \kappa_1 & = & y_n c_n + y_2 \frac{\left\langle l_1 p_2 \right\rangle}{\left\langle l_1 l_n \right\rangle},
 \nonumber \\
 \kappa_1 \lambda_1 & = & - y_n c_n \alpha_1 \lambda_n + y_2 \frac{\left\langle p_2 l_n \right\rangle}{\left\langle l_1 l_n \right\rangle}.
\eq
A possible solution is
\bq
 \kappa_1
 \;\; = \;\;
 1 + \frac{2 l_1 p_2}{2 l_1 l_n},
 & &
 \kappa_n 
 \;\; = \;\;
 \frac{c_1}{c_n},
 \\
 y_2 
 \;\; = \;\;
 \frac{\left[ l_1 p_2 \right]}{\left[ l_1 l_n \right]},
 & &
 y_n 
 \;\; = \;\;
 \frac{1}{c_n},
 \nonumber \\
 \lambda_1
 \;\; = \;\;
 -
 \frac{2 l_1 l_n  \left\langle l_1 + \left| p\!\!\!/_2 \right| l_n + \right\rangle}
      {\left(2 l_1 l_n\right)^2 - p_1^2 p_n^2 + \left(2 l_1 l_n \right) \left(2 l_1 p_2\right)},
 & &
 \lambda_n
 \;\; = \;\;
 \frac{p_1^2 \left\langle l_1 + \left| p\!\!\!/_2 \right| l_n + \right\rangle}
      {\left(2 l_1 l_n\right)^2 - p_1^2 p_n^2 + \left(2 l_1 l_n \right) \left(2 l_1 p_2\right)}.
 \nonumber
\eq

% ----------------------------------------------------------------------------------

\section{The large $z$-behaviour in the eikonal approximation}
\label{appendix:eikonal}

Let us consider a theory with massless or massive scalar ``hard'' particles, denoted by a hat 
and QCD-like ``soft'' particles (gluons, quarks, anti-quarks), denoted without a hat.
The momenta of the hard particles are of order $z^1$, the momenta of the soft particles are of order $z^0$.
The Feynman rules for this toy theory are as follows:
The hard particles interact only through three-valent vertices.
The Feynman rule for the three-valent vertex involving three hard particles with the cyclic order
$(\hat{1},\hat{2},\hat{3})$ is simply $i$, for the cyclic order
$(\hat{1},\hat{3},\hat{2})$ we have $(-i)$.
Furthermore there is a three-valent vertex, involving two hard particles and one soft gluon.
The Feynman rule for the cyclic order $(\hat{1},2,\hat{3})$ reads
\bq
 i \left(\hat{p}_1^\mu- \hat{p}_3^\mu \right).
\eq
There are no vertices involving only one hard particle.
The Feynman rules for the vertices involving only soft particles are the standard (cyclic-ordered) QCD Feynman rules,
listed in appendix~\ref{appendix:cyclic_ordered_rules}.

Let us consider the situation of three hard particles $\hat{1}$, $\hat{2}$ and $\hat{n}$ and $(n-3)$ soft particles
$3$, ..., $(n-1)$.
We assume particle $\hat{2}$ to be massless and particles $\hat{1}$ and $\hat{n}$ to have the same mass $m$ (which may be zero).
We will denote an amplitude in this toy theory by
\bq
   A_n^{\mathrm{eikonal}}\left(\hat{1},3,...,i,\hat{2},i+1,...,n-1,\hat{n}\right),
\eq
and we define
\bq
\label{I_n_eikonal}
 I_n^{\mathrm{eikonal}}\left(z\right)
 & = &
 \sum\limits_{i=2}^{n-1}
  \left( 2 \hat{p}_2 \hat{p}_n + \sum\limits_{j=i+1}^{n-1} 2 \hat{p}_2 p_j \right)
  A_n^{\mathrm{eikonal}}\left(\hat{1},3,...,i,\hat{2},i+1,...,n-1,\hat{n}\right).
\eq
We would like to show that $I_n^{\mathrm{eikonal}}(z)$ goes to a constant for large $z$.
Then the quantity 
\bq
 O_3 \; I_n^{\mathrm{eikonal}}(z)
\eq
with $O_3$ defined as in eq.~(\ref{O_3}) falls off like $1/z$.

It will be convenient to introduce soft currents
\bq
 J^\mu_{\mathrm{soft}}\left(a,...,b\right),
\eq
involving $(b-a+1)$ soft on-shell particles $a, a+1, ..., b$ and one soft off-shell gluon leg.
The momentum of this soft current is
\bq
 P & = & \sum\limits_{k=a}^b p_k.
\eq
We may group the Feynman diagrams contributing to $I_n^{\mathrm{eikonal}}(z)$
into sets, where exactly $r$ soft currents couple to the hard particles $\hat{1}$, $\hat{2}$
and $\hat{n}$ with $1 \le r \le n-3$.
Therefore we have a decomposition
\bq
 I_n^{\mathrm{eikonal}}(z)
 & = & 
 \sum\limits_{r=1}^{n-3}
 I_{n,r}^{\mathrm{eikonal}}(z).
\eq
We will show that each contribution $I_{n,r}^{\mathrm{eikonal}}(z)$ individually goes to a constant for large $z$.

Let us discuss $I_{n,r}^{\mathrm{eikonal}}(z)$ with $r$ soft currents $J^{\mathrm{soft}}_1$, ..., $J^{\mathrm{soft}}_r$
and associated momenta $P_1$, ..., $P_r$.
The cyclic order among the soft currents is respected in each diagram contributing to $I_{n,r}^{\mathrm{eikonal}}(z)$.
We will use the notation
\bq
 P_{a,a+1,...,b} & = &
 \sum\limits_{k=a}^b P_k.
\eq
Let us first discuss the situation, where two or more soft currents couple to the hard line $\hat{2}$.
These contributions add up to zero in $I_{n,r}^{\mathrm{eikonal}}(z)$.
In order to see this, consider the situation, where the two outermost soft currents coupling to $\hat{2}$ are 
$J^{\mathrm{soft}}_a$ and $J^{\mathrm{soft}}_{a+1}$.
\begin{figure}
\begin{center}
\includegraphics[bb= 80 632 525 745]{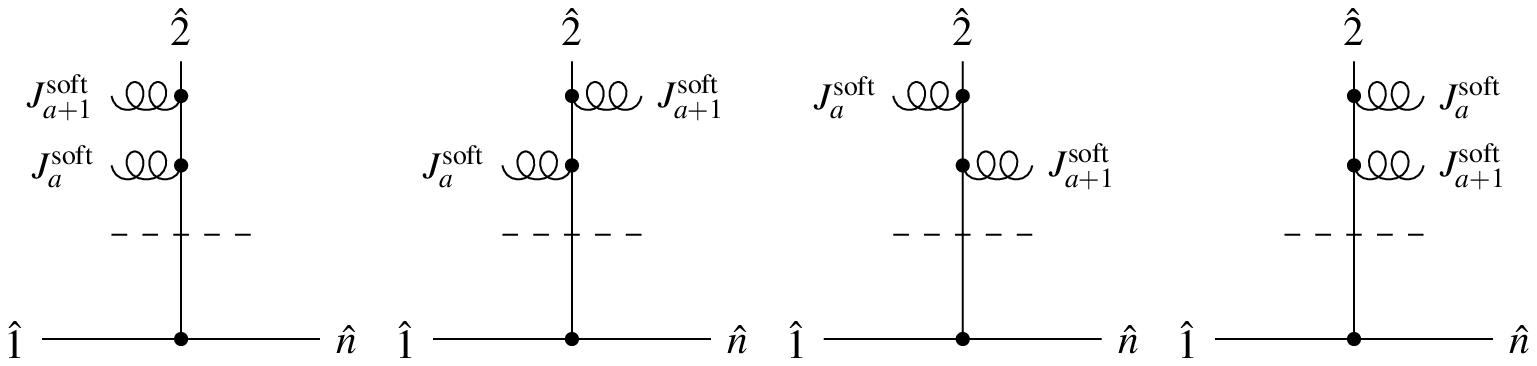}
\caption{\label{fig1}
Diagrams showing how the two outermost soft currents $J^{\mathrm{soft}}_a$ and $J^{\mathrm{soft}}_{a+1}$ may couple to the hard particle $\hat{2}$.
There may be further soft currents attached to the hard lines below the dashed line.
}
\end{center}
\end{figure}
There are four possibilites, how these soft currents may couple to $\hat{2}$, shown in fig.~(\ref{fig1}).
With the inclusion of the prefactors of the fundamental BCJ-relation, these contributions add up to zero.

Let us now consider the case, where one soft current $J^{\mathrm{soft}}_a$ couples to the hard particle $\hat{2}$.
Here we get the contribution
\bq
 - \left( 2 \hat{p}_2 P_a \right)  \frac{2 \hat{p}_2 J^{\mathrm{soft}}_a}{\left(\hat{p}_2+P_a\right)^2}
 & = &
 2 \hat{p}_1 J^{\mathrm{soft}}_a
 +
 2 \hat{p}_n J^{\mathrm{soft}}_a
 +
 {\mathcal O}\left(z^0\right).
\eq
We may now add up all contributions and obtain
\bq
 I_{n,r}^{\mathrm{eikonal}}(z)
 & = & 
 \sum\limits_{i=0}^r \left( 2 \hat{p}_2 \hat{p}_n + \sum\limits_{j=i+1}^{r} 2 \hat{p}_2 P_j \right)
 \left(-1\right)^i
 \left( \prod\limits_{k=1}^i \frac{2 \hat{p}_1 J^{\mathrm{soft}}_k}{2 \hat{p}_1 P_{1,...,k} } \right)
 \left( \prod\limits_{l=i+1}^r \frac{2 \hat{p}_n J^{\mathrm{soft}}_l}{2 \hat{p}_n P_{l,...,r} } \right)
 \nonumber \\
 & &
 + \sum\limits_{i=1}^r
 \left(-1\right)^{i-1}
 \left( \prod\limits_{k=1}^{i-1} \frac{2 \hat{p}_1 J^{\mathrm{soft}}_k}{2 \hat{p}_1 P_{1,...,k} } \right)
 \left( \prod\limits_{l=i+1}^r \frac{2 \hat{p}_n J^{\mathrm{soft}}_l}{2 \hat{p}_n P_{l,...,r} } \right)
 \left(  2 \hat{p}_1 J^{\mathrm{soft}}_i + 2 \hat{p}_n J^{\mathrm{soft}}_i \right)
 \nonumber \\
 & &
 +
 {\mathcal O}\left(z^0\right).
\eq
The terms in the first sum come from diagrams, where all soft currents couple either to the hard particle $\hat{1}$ or $\hat{n}$,
the terms of the second sum correspond to diagrams, where exactly one soft current couples to the hard particle $\hat{2}$. 
Noting that
\bq
 2 \hat{p}_2 \hat{p}_n + \sum\limits_{j=i+1}^{r} 2 \hat{p}_2 P_j 
 & = &
 2 \hat{p}_1 P_{1,...,i} - 2 \hat{p}_n P_{i+1,...,r}
 +
 {\mathcal O}\left(z^0\right)
\eq
one sees that
\bq
 I_{n,r}^{\mathrm{eikonal}}(z)
 & = & 
 {\mathcal O}\left(z^0\right),
\eq
as claimed.

\section{Cyclic-ordered Feynman rules}
\label{appendix:cyclic_ordered_rules}

In this appendix we give a list of the cyclic-ordered Feynman rules. 
They are obtained from the standard Feynman rules 
by extracting from each formula the coupling constant and the colour part.
The propagators for quark and gluon particles are given by
\bq
\begin{picture}(85,20)(0,5)
 \ArrowLine(70,10)(20,10)
\end{picture} 
 & = &
 i\frac{p\!\!\!/+m}{p^2-m^2},
 \nonumber \\
\begin{picture}(85,20)(0,5)
 \Gluon(20,10)(70,10){-5}{5}
\end{picture} 
& = &
 \frac{-ig^{\mu\nu}}{p^2}.
\eq
The cyclic-ordered Feynman rules for the three-gluon and the four-gluon vertices are
\bq
\begin{picture}(100,35)(0,50)
\Vertex(50,50){2}
\Gluon(50,50)(50,80){3}{4}
\Gluon(50,50)(76,35){3}{4}
\Gluon(50,50)(24,35){3}{4}
\LongArrow(56,70)(56,80)
\LongArrow(67,47)(76,42)
\LongArrow(33,47)(24,42)
\Text(60,80)[lt]{$p_1^{\mu_1}$}
\Text(78,35)[lc]{$p_2^{\mu_2}$}
\Text(22,35)[rc]{$p_3^{\mu_3}$}
\end{picture}
 & = &
 i \left[ g^{\mu_1\mu_2} \left( p_1^{\mu_3} - p_2^{\mu_3} \right)
         +g^{\mu_2\mu_3} \left( p_2^{\mu_1} - p_3^{\mu_1} \right)
         +g^{\mu_3\mu_1} \left( p_3^{\mu_2} - p_1^{\mu_2} \right)
   \right],
 \nonumber \\
 \nonumber \\
 \nonumber \\
\begin{picture}(100,35)(0,50)
\Vertex(50,50){2}
\Gluon(50,50)(71,71){3}{4}
\Gluon(50,50)(71,29){3}{4}
\Gluon(50,50)(29,29){3}{4}
\Gluon(50,50)(29,71){3}{4}
\Text(72,72)[lb]{\small $\mu_1$}
\Text(72,28)[lt]{\small $\mu_2$}
\Text(28,28)[rt]{\small $\mu_3$}
\Text(28,72)[rb]{\small $\mu_4$}
\end{picture}
 & = &
  i \left[
        2 g^{\mu_1\mu_3} g^{\mu_2\mu_4} - g^{\mu_1\mu_2} g^{\mu_3\mu_4} 
                                        - g^{\mu_1\mu_4} g^{\mu_2\mu_3}
 \right].
 \nonumber \\
 \nonumber \\
\eq
The Feynman rule for the quark-gluon vertex is given by
\bq
\label{colour_ordered_quark_gluon_antiquark_vertex}
\begin{picture}(100,35)(0,50)
\Vertex(50,50){2}
\Gluon(50,50)(80,50){3}{4}
\ArrowLine(50,50)(29,71)
\ArrowLine(29,29)(50,50)
\Text(82,50)[lc]{$\mu$}
\end{picture}
 \;\; = \;\;
 i \gamma^{\mu},
 & \;\;\;\;\;\;\;\;\; &
\begin{picture}(100,35)(0,50)
\Vertex(50,50){2}
\Gluon(50,50)(20,50){3}{4}
\ArrowLine(50,50)(71,71)
\ArrowLine(71,29)(50,50)
\Text(18,50)[rc]{$\mu$}
\end{picture}
 \;\; = \;\;
 -i \gamma^{\mu}.
 \nonumber \\
 \nonumber \\
\eq

\end{appendix}

% ----------------------------------------------------------------------------------
% references
\bibliography{/home/stefanw/notes/biblio}
\bibliographystyle{/home/stefanw/latex-style/h-physrev5}

\end{document}